\documentclass[fleqn,preprintnumbers,amsmath,amssymb,prb]{article}
\usepackage{graphicx}
\usepackage{dcolumn}
\usepackage{bm}
\newcommand{\newc}{\newcommand}
\newc{\beq}{\begin{equation}}
\newc{\eeq}{\end{equation}}
\newc{\beqa}{\begin{eqnarray*}}
\newc{\eeqar}{\end{eqnarray}}
\newc{\beqar}{\begin{eqnarray}}
\newc{\eeqa}{\end{eqnarray*}}
\newc{\bd}{\begin{displaymath}}
\newc{\ed}{\end{displaymath}}
\newc{\mbf}{\mathbf}

\begin{document}
\title{Eigenvalues of the Anti-periodic Calogero - Sutherland Model}
\author{Arindam Chakraborty, Subhankar Ray 
\thanks{Email: subho@juphys.ernet.in} \\
Dept. of Physics, Jadavpur University, Kolkata 700032, India}

\date{5 October 2004}

\maketitle
\begin{abstract}
The $U(1)$ Calogero Sutherland Model (CSM) with anti-periodic
boundary condition is studied. The Hamiltonian is reduced to
a convenient form by similarity transformation.
The matrix representation of the Hamiltonian acting on a 
partially ordered state space is obtained in an upper triangular
form. Consequently the diagonal elements become the energy 
eigenvalues. 

\end{abstract}
\setcounter{equation}{0}
\setcounter{page}{1}

\section{Introduction}

In recent years many exactly solvable models have been proposed
for quantum many body systems in one dimension. The basic
advantage of dealing with such one dimensional problem is that 
due to highly restrictive spatial degrees of freedom, there
exist several algebraic techniques that are applicable.
The well studied Bethe Ansatz technique is used to solve models
with nearest neighbour exchange interaction \cite{bethe31}.
But for long range interaction a few of the recently proposed 
models prove to have direct correspondence with the physical
reality.

Recently there have been much interests in the class of models 
with the interaction that falls off as a inverse square of the 
distance between a pair of particles or spins. 
The study of exact features of long-range interaction in this type
of model dates back to Moser \cite{moser75}, Calogero
\cite{calo62}, Sutherland \cite{suthJMP,suthPRA}.
The Hamiltonian of the Calogero Sutherland Model (CSM)
in units of $\hbar^2/2m$ is given by,
\beq
H = - \sum_{j=1}^N \frac{\partial^2}{\partial x^2} + \sum_{j<k} 
\frac{2 \alpha ( \alpha-1)}{d^2 (x_j-x_k)}
\eeq
where $\alpha$ is the interaction parameter and
$d(x)$ is the length between sites,
\bd
d(x) = \frac{L}{\pi} \sin \left( \frac{\pi x}{L} \right)
\ed
$L$ is the total length of the one dimensional chain.

One of the most important practical features of the model is
that the inverse square potential can be regarded as a pure
statistical interaction and the model maps to an ideal gas
of particle obeying the fractional statistics 
\cite{ha94,ha95,hald93}.

The eigenstate of $U(1)$ CSM can be written in terms of Jack
symmetric polynomial \cite{hald93,forr94} whose algebraic properties
provide a powerful and direct method of calculating the
most general correlation function.
The exact calculation of the correlation function further
provide conclusive evidences of the inherent
fractional exclusion and exchange statistics embodied
in CSM \cite{forr93,forr94}. 
The CSM is directly related to several interesting physical and
mathematical problems, e.g., Selberg integral \cite{forr93,forr94},
$W^{\infty}$ algebra \cite{hikwad}, edge states of quantum Hall
droplet \cite{hald93,ha95}, random matrix theory \cite{dyson62,meheta}, 
Jack symmetric polynomials \cite{jack}.

In our present paper we have calculated 
the energy eigenvalues of the CSM Hamiltonian with
anti-periodic boundary condition. Such boundary condition
is relevant where a magnetic field pierces perpendicular
to the plane of one dimensional ring
(Topological representation of one dimensional linear chain
of sites). When a particle goes around the entire system $n$
times it picks up a net phase $\exp(i n \phi)$.
Consequently the interaction term takes the following 
form \cite{habook},
\beqar
\sum _{n=- \infty}^{+\infty}\frac{\exp{(i\phi n)}}{(x+nL)^2}
\eeqar
where $L =$ length of the chain.
Let $\phi=2\pi p/q$ and $n=jq+k$, where $p$ and $q$ are 
mutual primes and $j$ and $k$ are integer such that 
\bd
-\infty < j < +\infty, \hskip 2cm 0 \le k \le q-1. 
\ed
Therefore, the sum in the above expression, Eq.(2)
can be rewritten as,
\beq\label{eq3}
\sum _{k=0}^{q-1} \sum _{j=- \infty}^{+\infty}
\frac{1}{\left[(x+kL)+(qL)j \right] ^2}   
=\sum _{k=1}^{q-1}\frac{\exp{(i 2\pi px/q)}}
{\left[(qL/\pi) \sin{\frac{\pi(x+kL)}{qL}}\right]^2}
\eeq
The effective length of the chain is thus ($q L$).
The above interaction term represents in general the
so called twisted boundary condition \cite{habook}.
When $p/q = 1/2$ the corresponding particle interaction 
term becomes,
\beq\label{eq4}
\frac{\cos{(\pi x/L)}}{\frac{L^2}{\pi^2}\sin^2{(\pi x/L)}}
\eeq
Eq.(\ref{eq4}) gives the interaction term with
anti-periodic boundary condition. Under such 
boundary condition the final expression of the model
Hamiltonian becomes,
\beq\label{eq5}
H=-\sum _{j=1}^{N}\frac{\partial ^2}{\partial {x_j}^2}+2 \alpha (\alpha-1)
\frac{\pi ^2}{L^2}\sum _{j<k}\frac{\cos{[(\pi/L)(x_j-x_k)]}}{\sin^2{[(\pi/L)
(x_j-x_k)]}}
\eeq
The diagonalization of such Hamiltonian is usually done
after reducing it to a convenient form by means of
successive similarity transformations.
Here the final form of the Hamiltonian is diagonalized with
the help of the symmetric polynomial type eigenfunctions,
known as the Jack symmetric polynomials.
The matrix representation of the Hamiltonian becomes
triangular on a partially ordered state space.
The action of the Hamiltonian on such a state space
gives rise to several {\it mother} and {\it daughter}
states. The mother and daughter states are connected by a 
particular rule. And a topological representation of
the excitation spectrum is obtained for a specific example.

\section{Simplification of the model Hamiltonian}
\subsection{Similarity transformation}

Let us put ($\lambda=-\alpha$) in Eq.(5).
The the expression for the Hamiltonian becomes,
\beq\label{eq6}
H=-\sum _{j=1}^{N}\frac{\partial ^2}{\partial {x_j}^2}+2 \lambda
(\lambda+1)
\frac{\pi ^2}{L^2}\sum _{j<k}\frac{\cos{[(\pi/L)(x_j-x_k)]}}{\sin^2{[(\pi/L)
(x_j-x_k)]}}
\eeq
Eq.(6) can be rewritten in the following form, apart from some
constant terms.
\begin{eqnarray}\label{eq7}
H= - \sum _{j=1}^{N}\frac{\partial ^2}{\partial {x_j}^2}
+ 4 \lambda \frac{\pi^2}{L^2}
\sum _{j<k}\frac{\cos{[\frac{\pi}{L}x_{jk}]}}{\sin^2{[\frac{\pi}{L}
x_{jk}]}}+2 \lambda (1-\lambda)\frac{\pi^2}{L^2}\sum _{j<k}
\frac{\cos[\frac{\pi}{L}x_{jk}]+1}{\sin^2[\frac{\pi}{L}x_{jk}]} && \nonumber \\
- 2 \lambda (1-\lambda) \frac{\pi^2}{L^2}\sum _{j<k} \left(\frac{1}
{\sin^2{[\frac{\pi}{L}x_{jk}]}}-1\right)+4 \lambda 
(\lambda-1)\frac{\pi^2}{L^2} \sum _{j<k}\frac{
\cos[\frac{\pi}{L}x_{jk}]}{\sin^2{[\frac{\pi}{L}x_{jk}]}} &&
\end{eqnarray}
where $x_{jk} = x_j - x_k$.

Putting $\omega_j=\exp(i\pi x_j/L)$, where $i=\sqrt{-1}$, Eq.(7)
becomes,
\begin{eqnarray}\label{eq8}
H=\frac{\pi^2}{L^2}\left[\sum_{j=1}^N(\omega_j\frac{\partial}{\partial 
\omega_j})^2+ 4 \lambda\sum_{j<k}\left(\omega_j\frac{\partial}
{\partial\omega_j}-\omega_k\frac{\partial}{\partial\omega_k}\right)
\frac{\omega_j\omega_k}{\omega^2_j-\omega^2_k} \right.\\ \nonumber 
-4 \lambda (1-\lambda)\sum_{j<k}\frac{\omega_j\omega_k}
{(\omega_j-\omega_k)^2}
+2 \lambda(1-\lambda)\sum_{j<k}\left(\frac{\omega_j^2+\omega_k^2}
{\omega_j^2
-\omega_k^2}\right)^2 \\\nonumber
\left. -8 \lambda(\lambda-1)\sum_{j<k}\frac{\omega_j^2+\omega_k^2}
{(\omega_j^2-\omega_k^2)^2}\;\; \omega_j\omega_k \right]
\end{eqnarray}
Let's take a similarity transformation with the following ansatz,
\begin{eqnarray}\label{eq9}
\psi=\prod_{j<k}\left(\frac{\omega_j}{\omega_k}-\frac{\omega_k}{\omega_j}\right)^\lambda \phi
\end{eqnarray}
Therefore, the Hamiltonian in units of $\pi^2/L^2$ becomes,
\begin{eqnarray}\label{eq10}
\hat{H}=\sum_{j=1}^{N}\left[\omega_j\frac{\partial}{\partial\omega_j}\right]^2
+2\lambda\sum_{j<k}\left(\frac{\omega_j+\omega_k}{\omega_j-\omega_k}\right)
\left[ \omega_i\frac{\partial}{\partial \omega_j}-\omega_k \frac{\partial}
{\partial \omega_k} \right] \\\nonumber
+4\lambda(\lambda-1)\sum_{j<k}\frac{\omega_j\omega_k}{(\omega_j-\omega_k)^2}
\end{eqnarray}
Again using the similarity
\beq\label{eq11}
\phi=\prod_{j<k}\left(\frac{\omega_j}{\omega_k}+\frac{\omega_k}
{\omega_j}-2\right)^{\beta/2}\phi_0
\eeq
where $\beta=\beta(\lambda)$.
The above Hamiltonian takes the form 
\begin{eqnarray}\label{eq12}
\tilde{H}=\sum_{j=1}^{N}\left[\omega_j\frac{\partial}{\partial\omega_j}\right]^2
+ (2\lambda+\frac{3}{2}\beta(\lambda))\sum_{j<k}\left(\frac{\omega_j+\omega_k}{\omega_j-\omega_k}\right)
\left[ \omega_i\frac{\partial}{\partial \omega_j}-\omega_k \frac{\partial}
{\partial \omega_k} \right] 
\end{eqnarray}
where 
\begin{eqnarray}\label{eq13}
\beta(\lambda)=\frac{1}{2}\left(-1\pm \sqrt{1+8\lambda-8\lambda^2}\right)
\end{eqnarray}
Introducing $A(\lambda)=2\lambda+(3/2)\beta(\lambda)$, we get
the following expression for the model-Hamiltonian,
\beqar\label{eq14}
\tilde{H}=\sum_{j=1}^{N}\left[\omega_j\frac{\partial}{\partial\omega_j}\right]^2
+A(\lambda)\sum_{j<k}\left(\frac{\omega_j+\omega_k}{\omega_j-\omega_k}\right)
\left[ \omega_i\frac{\partial}{\partial \omega_j}-\omega_k \frac{\partial}
{\partial \omega_k} \right] 
\eeqar
Hence $\tilde{H}$ has two additive parts, one representing free particle 
Hamiltonian, and a second part containing the reduced interaction term.
\beq\label{eq15}
\tilde{H}=H_0+A(\lambda)H_1
\eeq

\subsection{Action of the Hamiltonian on ordered states}
The eigenstates of the present Hamiltonian,
Eq.(\ref{eq15}), can be given by 
the following Bosonic state,
\begin{eqnarray}\label{eq16}
\vert n_1....n_N\rangle=\sum_P\prod_{j=1}^{N}\omega_j^{n_{_{P_j}}}
\end{eqnarray}
The set $\{n_i \vert i =1...N\}$
can be considered as a Bosonic 
quantum number with no restriction on their values.
Without loss of generality we can introduce an ordering
($n_1\ge n_2 \ge n_3 \ge ....\ge n_N$). The action of 
$H_0$ and $H_1$ on such an ordered state gives,
\begin{eqnarray}\label{eq17}
H_0\vert n_1....n_N\rangle=\left(\sum_{j=1}^{N}n_j^2\right)\vert n_1....n_N\rangle
\end{eqnarray}
and
\beq\label{eq18}
H_1\vert n_1....n_N\rangle=\sum_{j<k}(n_j-n_k)\left(\vert n_1....n_N\rangle 
+2 \sum_{p=1}^{n_j-n_k-1}\vert,\dots,n_j-p,\dots,n_k+p,\dots\rangle \right)
\eeq

\section{The Eigenvalues of the Hamiltonian}
\subsection{Level of states} 
We define $\vert n_1...n_N\rangle$ as the mother state 
and the states generated by squeezing a pair of quantum 
number by one unit like $\vert\dots,n_j-1,\dots,n_k+1,\dots\rangle$ 
as the daughter state.
Squeezing of mother state into daughter states 
is permitted when such squeezing retains ordering of
$n_j$s, i.e, when ($n_j-n_k\ge 2$).

The family of states can be organized into levels such that the 
members of a given level are mutually unrelated in a sense that 
they are unreachable from each other.
The daughter of a member from any given level belongs to a lower 
level in the family.
Let, $\vert u\rangle_1$ implies the highest level mother state.
The state index $u$ represents the total number of levels in the family.
$\vert u\rangle_{\mu}$ represents the family of daughter state with 
$1\le u'\le u$ and $\mu=$ is an index for the state in the $u$th 
level.

\subsection{Generation of Mother and Daughter states}

Let us take $\vert6,4,3,1 \rangle$ as the mother state.
It is obvious that by squeezing a relevant pair of
quantum numbers we obtain a sequence of mother and daughter states.
Finally we reach an irreducible daughter state that cannot 
serve as a mother in order to produce daughter state(s).
This irreducible state is called the ground state, and is 
commonly denoted by $\vert1\rangle_1$.
Table.I shows the possible mother and daughter states
starting with $\vert6,4,3,1 \rangle$ as the highest level
mother state. \\

Table.I: Mother and daughter state association. \\
\vskip .3cm
\begin{tabular}{c c}
\begin{tabular}{|c|c|}  \hline \hline
Mother State & Daughter State \\ \hline \hline
$\vert 6,4,3,1 \rangle = \vert6 \rangle_1$ & 
$\vert5,5,3,1\rangle = \vert5 \rangle_1$ \\
& $\vert6,4,2,2 \rangle = \vert5 \rangle_2$ \\
& $\vert5,4,4,1 \rangle = \vert4 \rangle_1$ \\
& $\vert6,3,3,2 \rangle = \vert4 \rangle_2$ \\
& $\vert5,4,3,2 \rangle = \vert3 \rangle_1$ \\ \hline
$\vert5,5,3,1 \rangle = \vert5\rangle_1$ &
$\vert5,4,4,1\rangle = \vert4\rangle_1$ \\
&$\vert5,5,2,2\rangle = \vert4\rangle_3$ \\
&$\vert5,4,3,2\rangle = \vert3\rangle_1$ \\ \hline
$\vert6,4,2,2\rangle = \vert5\rangle_2$ &
$\vert6,3,3,2\rangle = \vert4\rangle_2$ \\
&$\vert5,5,2,2\rangle = \vert4\rangle_3$ \\
&$\vert5,4,3,2\rangle = \vert3\rangle_1$ \\ \hline
$\vert5,4,4,1\rangle = \vert4\rangle_1$ &
$\vert5,4,3,2\rangle = \vert3\rangle_1$ \\
&$\vert4,4,4,2\rangle = \vert2\rangle_1$ \\ \hline
\end{tabular}
&
\begin{tabular}{|c|c|}  \hline \hline
Mother State & Daughter State \\ \hline \hline
$\vert6,3,3,2\rangle = \vert4\rangle_2$ &
$\vert5,4,3,2\rangle = \vert3\rangle_1$ \\
&$\vert5,3,3,3\rangle = \vert2\rangle_2$ \\ \hline
$\vert5,5,2,2\rangle = \vert4\rangle_3$ &
$\vert5,4,3,2\rangle = \vert3\rangle_1$ \\ \hline 
$\vert5,4,3,2\rangle = \vert3\rangle_1$ &
$\vert4,4,4,2\rangle = \vert2\rangle_1$ \\
&$\vert5,3,3,3\rangle = \vert2\rangle_2$ \\ 
&$\vert4,4,3,3\rangle = \vert1\rangle_1$ \\ 
& (irreducible daughter \\
& state) \\ \hline
$\vert4,4,4,2\rangle = \vert2\rangle_1$ &
$\vert4,4,3,3\rangle = \vert1\rangle_1$ \\
& (irreducible daughter \\
& state) \\ \hline
$\vert5,3,3,3\rangle = \vert2\rangle_2$ &
$\vert4,4,3,3\rangle = \vert1\rangle_1$ \\
& (irreducible daughter \\
& state) \\ \hline
\end{tabular}
\end{tabular}

\vspace{.6cm}

Moreover as the states in the same level are not connected,
the Hamiltonian is diagonal in that subspace. The following topological
representation shows the connection of mother and daughter state
in the excitation spectrum.
%
%
\begin{figure}[h]
\resizebox{!}{3.0in}
{\hskip 5cm \includegraphics{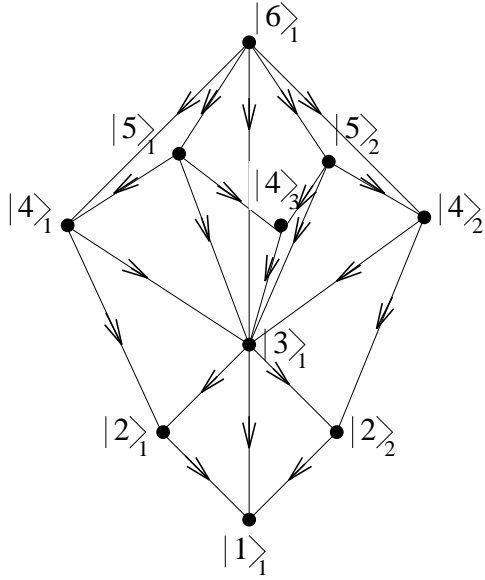}}
\caption{The topological representation of the excitation spectrum}
\end{figure}
Let us define $\nu(\eta)$ as the multiplicity of a number in
a given state ket taken as a mother state. The weight of an arrow
$W$ is given by the following equation,
\beq\label{eq19}
W = \nu(\eta_i) \nu(\eta_j)
\eeq

The following table gives the respective weights of the possible
transitions from the mother to daughter states.

%
%
Table.II: Transitions and weights for Calogero-Sutherland model
with anti-periodic boundary condition \\

\begin{tabular}{c c}
\begin{tabular}{|c|c|}  \hline \hline
Transitions & Weights (W) \\ \hline \hline
$\vert 6\rangle_1 = \vert5 \rangle_1$ & $1 \cdot 1 = 1$ \\
$\vert 6\rangle_1 = \vert5 \rangle_2$ & $1 \cdot 1 = 1$ \\
$\vert 6\rangle_1 = \vert4 \rangle_1$ & $1 \cdot 1 = 1$ \\
$\vert 6\rangle_1 = \vert4 \rangle_2$ & $1 \cdot 1 = 1$ \\
$\vert 6\rangle_1 = \vert3 \rangle_1$ & $1 \cdot 1 = 1$ \\ \hline
$\vert 5\rangle_1 = \vert4 \rangle_1$ & $2 \cdot 1 = 2$ \\
$\vert 5\rangle_1 = \vert4 \rangle_3$ & $1 \cdot 1 = 1$ \\
$\vert 5\rangle_1 = \vert3 \rangle_1$ & $2 \cdot 1 = 2$ \\ \hline
$\vert 5\rangle_2 = \vert4 \rangle_2$ & $1 \cdot 2 = 2$ \\
$\vert 5\rangle_2 = \vert4 \rangle_3$ & $1 \cdot 1 = 1$ \\
$\vert 5\rangle_2 = \vert3 \rangle_1$ & $1 \cdot 2 = 2$ \\ \hline
\end{tabular}
&
\begin{tabular}{|c|c|}  \hline \hline
Transitions & Weights (W) \\ \hline \hline
$\vert 4\rangle_1 = \vert3 \rangle_1$ & $2 \cdot 1 = 2$ \\
$\vert 4\rangle_1 = \vert2 \rangle_1$ & $1 \cdot 1 = 1$ \\ \hline
$\vert 4\rangle_2 = \vert3 \rangle_1$ & $1 \cdot 2 = 2$ \\
$\vert 4\rangle_2 = \vert2 \rangle_2$ & $1 \cdot 1 = 1$ \\ \hline
$\vert 4\rangle_3 = \vert3 \rangle_1$ & $2 \cdot 2 = 4$ \\ \hline
$\vert 3\rangle_1 = \vert2 \rangle_1$ & $1 \cdot 1 = 1$ \\
$\vert 3\rangle_1 = \vert2 \rangle_2$ & $1 \cdot 1 = 1$ \\
$\vert 3\rangle_1 = \vert1 \rangle_1$ & $1 \cdot 1 = 1$ \\ \hline
$\vert 2\rangle_1 = \vert1 \rangle_1$ & $3 \cdot 1 = 3$ \\ \hline
$\vert 2\rangle_2 = \vert1 \rangle_1$ & $1 \cdot 3 = 3$ \\ \hline
\end{tabular}
\end{tabular}

\subsection{Sub-family of states}
Let us introduce the so called sub-family of states which 
consists of the highest level mother state and all 
her reachable daughter states.
The total number of sub-family is 
called the dimension of the family.
Since the action of the Hamiltonian on a given state
space generates states belonging to lower levels, the
matrix representation of the Hamiltonian in a partially
ordered state space is always triangular. So the diagonal
terms of the Hamiltonian matrix are the energy eigenvalues.

The energy of an eigen-state spanned by a family with 
the highest level mother state $\vert n_1...n_N\rangle$ 
is given by 
\beq\label{eq20}
E^0(n_1,\dots,n_N)=\sum_{j=1}^N n_j^2 + A(\lambda)\sum_{j<k}(n_j-n_k)
\eeq
The off diagonal elements are found to be 
\beq\label{eq21}
{_{\mu'}} \langle u'\vert \tilde{H} \vert u\rangle_{_\mu}= \left\{
\begin{array}{l}
\sum_{P} \left(\prod_{i\in P}W_i\right) E^1(n_1,\dots,n_N) \hfill
\hskip 1cm \forall \mu >\mu' \\
\hfill \\
0 \hfill \forall \mu < \mu'
\end{array}
\right.
\eeq
with 
\beq\label{eq22}
E^1(n_1,\dots,n_N)=2A(\lambda)\sum_{j<k}(n_j-n_k)
\eeq
The sum in Eq.(21) is over all possible paths $P$ 
from $\vert u\rangle_{\mu}$
to $\vert u'\rangle_{\mu'}$ and the product is over all weights 
$W_i$ of the intermediates arrows belonging to $P$.
Then the general matrix element of the Hamiltonian is given by,
\beq\label{eq23}
{_{\mu'}} \langle u'\vert \tilde{H} \vert u\rangle_{_\mu} = 
\varepsilon_{u',\mu'}^{u,\mu}
\eeq

\section{Conclusion}

In this article we obtain the energy eigen-value
and eigen-states of the CSM with anti-periodic boundary 
condition (APBC). The model Hamiltonian, Eq.(5), is reduced 
to a convenient from by means of similarity transformation.
Finally we obtain an upper triangular representation of
the Hamiltonian in terms of a partially ordered basis.
The eigenfunctions are chosen to be symmetric polynomials,
known as the Jack polynomial.

\section*{Acknowledgement}

AC wishes to acknowledge the Council of Scientific and
Industrial Research, India (CSIR) for fellowship support.

%

\end{document}